\documentclass[usenatbib,usegraphicx,a4paper]{mn2e}
\usepackage{graphicx}
\usepackage{url}
\usepackage{amsmath}
\usepackage[T1]{fontenc}
\usepackage{ae,aecompl}

\newcommand{\slfrac}[2]{\left.#1\middle/#2\right.}

\title[Inferring the mass of sub-millimetre galaxies through magnification]{Inferring the mass of sub-millimetre galaxies by exploiting their gravitational magnification of background galaxies}

\author[H.~Hildebrandt et al.]{H.~Hildebrandt,$^{1,2}$ 
  L.~van~Waerbeke,$^{1}$ 
  D.~Scott,$^{1}$ 
  M.~B\'ethermin,$^{3,4}$ 
  J.~Bock,$^{5,6}$ \newauthor
  D.~Clements,$^{7}$ 
  A.~Conley,$^{8}$ 
  A.~Cooray,$^{5,9}$ 
  J.S.~Dunlop,$^{10}$
  S.~Eales,$^{11}$ 
  T.~Erben,$^{2}$ \newauthor 
  D.~Farrah,$^{12}$ 
  A.~Franceschini,$^{13}$
  J.~Glenn,$^{8,14}$ 
  M.~Halpern,$^{1}$ 
  S.~Heinis,$^{15}$ 
  R.J.~Ivison,$^{16,17}$ \newauthor
  G.~Marsden,$^{1}$ 
  S.J.~Oliver,$^{18}$ 
  M.~J.~Page,$^{19}$ 
  I.~P\'erez-Fournon,$^{20,21}$
  A.~J.~Smith,$^{18}$ \newauthor 
  M.~Rowan-Robinson,$^{7}$ 
  I.~Valtchanov,$^{22}$ 
  R.F.J.~van~der~Burg,$^{23}$
  J.D.~Vieira,$^{5}$ \newauthor 
  M.~Viero,$^{5}$ 
  L.~Wang,$^{18}$
\\
$^{1}$University of British Columbia, Department of Physics and Astronomy, 6224 Agricultural Road, Vancouver, B.C. V6T 1Z1,\\\:\:\:Canada \\
$^{2}$Argelander-Institut f\"ur Astronomie, Auf dem H\"ugel 71, 53121 Bonn, Germany \\ 
$^{3}$Laboratoire AIM-Paris-Saclay, CEA/DSM/Irfu - CNRS - Universit\'e Paris Diderot, CE-Saclay, pt courrier 131, F-91191 Gif-sur-\\\:\:\:Yvette, France \\
$^{4}$Institut d'Astrophysique Spatiale (IAS), b\^atiment 121, Universit\'e Paris-Sud 11 and CNRS (UMR 8617), 91405 Orsay, France \\
$^{5}$California Institute of Technology, 1200 East California Boulevard, Pasadena, California 91125, USA \\ 
$^{6}$Jet Propulsion Laboratory, 4800 Oak Grove Drive, Pasadena, California 91109, USA \\
$^{7}$Astrophysics Group, Imperial College London, Blackett Laboratory, Prince Consort Road, London SW7 2AZ, UK \\ 
$^{8}$Department of Astrophysical and Planetary Sciences, CASA 389-UCB, University of Colorado, Boulder, Colorado 80309, USA \\ 
$^{9}$Department of Physics \& Astronomy, University of California, Irvine, California 92697, USA \\ 
$^{10}$SUPA, Institute for Astronomy, University of Edinburgh, Royal Observatory, Edinburgh EH9 3HJ \\ 
$^{11}$Cardiff School of Physics and Astronomy, Cardiff University, Queens Buildings, The Parade, Cardiff CF24 3AA, UK \\ 
$^{12}$Department of Physics, Virginia Tech, Blacksburg, VA, 24061, USA \\ 
$^{13}$Dipartimento di Astronomia, Universit\`a di Padova, Vicolo Osservatorio, 3, 35122 Padova, Italy \\ 
$^{14}$Center for Astrophysics and Space Astronomy, University of Colorado, Boulder,  CO, USA \\ 
$^{15}$Department of Physics and Astronomy, The Johns Hopkins University, 3400 North Charles Street, Baltimore, MD 21218, USA \\
$^{16}$Institute for Astronomy, University of Edinburgh, Royal Observatory, Blackford Hill, Edinburgh EH9 3HJ, UK \\ 
$^{17}$UK Astronomy Technology Centre, Royal Observatory, Blackford Hill, Edinburgh EH9 3HJ, UK \\
$^{18}$Astronomy Centre, Department of Physics \& Astronomy, University of Sussex, Brighton BN1 9QH, UK \\ 
$^{19}$Mullard Space Science Laboratory, University College London, Holmbury St Mary, Dorking, Surrey RH5 6NT, UK \\ 
$^{20}$Instituto de Astrof\'isica de Canarias (IAC), E-38200 La Laguna, Tenerife, Spain \\ 
$^{21}$Departamento de Astrof\'isica, Universidad de La Laguna (ULL), Tenerife, Spain \\ 
$^{22}$European Space Astronomy Centre, Herschel Science Centre, ESA, 28691 Villanueva de la Ca\~nada, Spain \\ 
$^{23}$Leiden Observatory, Leiden University, Niels Bohrweg 2, 2333 CA Leiden, The Netherlands
}
   
\date{Released 2011 Xxxxx XX}

\pagerange{\pageref{firstpage}--\pageref{lastpage}} \pubyear{2011}

\begin{document}
\label{firstpage}

\maketitle
\clearpage
\begin{abstract}
  Dust emission at sub-millimetre wavelengths allows us to trace the
  early phases of star formation in the Universe. In order to
  understand the physical processes involved in this mode of star
  formation, it is essential to gain knowledge about the dark matter
  structures -- most importantly their masses -- that sub-millimetre
  galaxies live in. Here we use the magnification effect of
  gravitational lensing to determine the average mass and dust content
  of sub-millimetre galaxies with 250\,$\mu$m flux densities of
  $S_{\rm 250}>15\,{\rm mJy}$ selected using data from the Herschel
  Multi-tiered Extragalactic Survey. The positions of hundreds of
  sub-millimetre foreground lenses are cross-correlated with the
  positions of background Lyman-break galaxies at $z\sim3-5$ selected
  using optical data from the Canada-France Hawaii Telescope Legacy
  Survey. We detect a cross-correlation signal at the 7-$\sigma$ level
  over a sky area of one square degree, with $\sim80\%$ of this signal
  being due to magnification, whereas the remaining $\sim20\%$ comes
  from dust extinction. Adopting some simple assumptions for the dark
  matter and dust profiles and the redshift distribution enables us to
  estimate the average mass of the halos hosting the sub-millimetre
  galaxies to be $\log_{10} \left[M_{200}/ M_\odot\right] =
  13.17^{+0.05}_{-0.08}{\rm (stat.)}$ and their average dust mass
  fraction (at radii of $>10\,$kpc) to be $M_{\rm
    dust}/M_{200}\approx6\times10^{-5}$. This supports the picture
  that sub-millimetre galaxies are dusty, forming stars at a high
  rate, reside in massive group-sized halos, and are a crucial phase
  in the assembly and evolution of structure in the Universe.
\end{abstract}


\section{Introduction}
\label{sec:intro}
In the current picture galaxies form in the centres of the
gravitational potential wells of dark matter halos, when gas cools and
star formation sets in \citep{1962ApJ...136..748E}. However, many of
the observed properties of galaxies -- e.g., scaling laws between
different observables like the \cite{1976ApJ...204..668F} relation,
the \cite{1977A&A....54..661T} relation, and the m-$\sigma$ relation
\citep{2000ApJ...539L...9F} -- remain to be explained by a complete
theory of galaxy formation and evolution. We know from observations in
the $z<1$ Universe that star formation efficiency, i.e., the fraction
of baryonic mass turned into stars, depends on the environment, most
importantly on the mass of the dark matter halo as shown by
\cite{2012ApJ...744..159L}. The bulk of star formation, however,
happened at earlier cosmic epochs \citep{1996MNRAS.283.1388M}, when
the Universe was less than half of its current age, which necessitates
studies at high redshift ($z\ga1$).

Young stars are still enveloped in clouds of gas and dust, and
therefore most of their radiation is absorbed by dust, and re-emitted
at far-infrared wavelengths. Since this is the dominant source of
emission at these wavelengths, the luminosity of a galaxy in the
far-infrared wavelength range is directly related to the amount of
young stars, and therefore the rate of star formation. Sub-millimetre
telescopes are sensitive to this radiation. Moreover, the detection of
this light is less sensitive to redshift than in the optical, due to
the fortunate shape of the spectral energy distributions of galaxies
in the sub-millimetre regime yielding samples of star-forming galaxies
over a wide redshift range. Establishing a picture of the dark matter
environment of these sub-millimetre galaxies through measuring their
total (i.e., baryonic plus dark matter) masses is hence a fundamental
ingredient for understanding the build-up of stellar mass over cosmic
time.

Gravitational lensing is the most direct method of measuring mass in
the distant Universe, irrespective of whether it consists of dark or
baryonic matter, and independent of any assumptions about its
dynamical state. Thus, it represents a powerful tool for measuring the
masses of dark matter structures, both using fewer astrophysical
assumptions and being complementary to, e.g., velocity dispersion
measurements. In many cases the lensing effect of an individual
deflector is too weak to be detected. In weak gravitational lensing
\citep[WL; see e.g.][]{2001PhR...340..291B} a statistical approach is
used, averaging the signals of many lenses and/or sources.

In order to detect the effects of WL by a lens population, a suitable
source population is needed that lies behind the lenses (from the
observer's point of view). The extended redshift distribution of
sub-millimetre galaxies means that most potential background galaxies
are spatially unresolved in even the best ground-based optical
data. Hence the traditional shear technique of WL, which requires
ellipticity estimates, cannot be used in this particular case, and one
has to turn to the magnification technique, which does not require
resolved sources \citep{2010ApJ...723L..13V}.

Here we measure the WL-magnification effect of a sub-millimetre galaxy
sample to estimate their average dark matter halo density profiles. As
background sources we use optically-selected Lyman-break galaxies
\citep[LBGs;][]{1996ApJ...462L..17S} which are located at even higher
redshifts. The sub-millimetre as well as the optical data are
described in Sect.~\ref{sec:data}. An outline of the magnification
technique is given in Sect.~\ref{sec:technique}. Results are presented
in Sect.~\ref{sec:results} and discussed in
Sect.~\ref{sec:discussion}. A summary and an outlook are given in
Sect.~\ref{sec:outlook}. We assume a WMAP7 cosmology throughout
\citep{2011ApJS..192...18K}: ($\Omega _{\rm M}$, $\Omega_\Lambda$,
$h$, $\sigma_8$) $=$ (0.27, 0.73, 0.70, 0.81).

\section{Data}
\label{sec:data}
\subsection{Sub-millimetre data}
The sub-millimetre galaxies are selected from the Herschel
Multi-tiered Extragalactic Survey
\citep[HerMES;][]{2010A&A...518L..21O,2012arXiv1203.2562H}, observed
with the Spectral and Photometric Imaging Receiver
\citep[SPIRE;][]{2010A&A...518L...3G} on board the \emph{Herschel
  Space Observatory} in the Cosmic Evolution Survey
\citep[COSMOS;][]{2007ApJS..172....1S} field. The catalogue
\footnote{available at the HeDaM on-line data base:
  \url{http://hedam.oamp.fr/HerMES/}} was constructed by the HerMES
team, using blind extraction at 250\,$\mu$m and a prior-based
extraction for the other two wave bands at 350\,$\mu$m and
500\,$\mu$m. Details can be found in \cite{2010MNRAS.409...48R}. Note
that these sub-millimetre galaxies constitute the lenses here and not
the sources as in \cite{2011MNRAS.414..596W}.

Unlike 'classical' sub-millimetre sources detected at longer
wavelength like, e.g., SCUBA-type \citep{1999MNRAS.303..659H}
galaxies, the galaxies detected by SPIRE at 250\,$\mu$m do not show a
strong negative $k$-correction
\citep{1991A&AS...89..285F,1993MNRAS.264..509B}. Hence their redshift
distribution does not extend to as high redshifts, which is beneficial
for the magnification measurement presented here, since it allows for
easier separation between the sub-millimetre lenses and the LBG
sources in redshift.

The beam size of Herschel at 250\,$\mu$m is ${\rm
  FWHM}\approx18''$. The centroids of the sub-millimetre galaxies in
each sky coordinate are known to $\Delta\alpha=\Delta\delta=0.6\,{\rm
  FWHM/SNR}$, where SNR is the signal-to-noise ratio of the detection
\citep{2007MNRAS.380..199I}. The mean SNR of our lenses is
$\overline{\rm SNR}=5.3$, which corresponds to centroid errors of
$\Delta\alpha=\Delta\delta=2\farcs0$. Additionally, the pointing
accuracy of Herschel is limited to $\approx2''$ as well
\citep{2010A&A...518L...1P}. We add these two contributions to the
centroid error in quadrature and account for this angular uncertainty
in the modelling (see Sect.~\ref{sec:theory}).

\subsection{Optical data}
The central part of the COSMOS field (1\,deg$^2$) was targeted with
the MegaCam imager on the Canada France Hawaii Telescope (CFHT) as
part of the CFHT Legacy Survey. Very deep images (5-$\sigma$ limiting
magnitudes for point sources of 26.5-28.0) in the $ugriz$-bands from
the CARS project \citep{2009A&A...493.1197E,2009A&A...498..725H} were
used to select a sample of $\sim17\,500$ Lyman-break galaxies
\citep{2009A&A...498..725H,2009A&A...507..683H} at redshifts
$z\sim3-5$.  These are too distant to be resolved in the optical data,
but they constitute an ideal sample for magnification measurements.

This set of LBGs has been studied in detail in the literature. Their
clustering properties are described in \cite{2009A&A...498..725H} and
a measurement of the LBG luminosity function based on this data set is
presented in \cite{2010A&A...523A..74V}. Furthermore, these LBGs were
already used as background sources for a magnification measurement in
a previous study \citep{2009A&A...507..683H}, but then using normal
galaxies selected by photometric redshifts as lenses.

\subsection{Redshift distributions}
\label{sec:z-dist}
\begin{figure}
\includegraphics[width=\hsize]{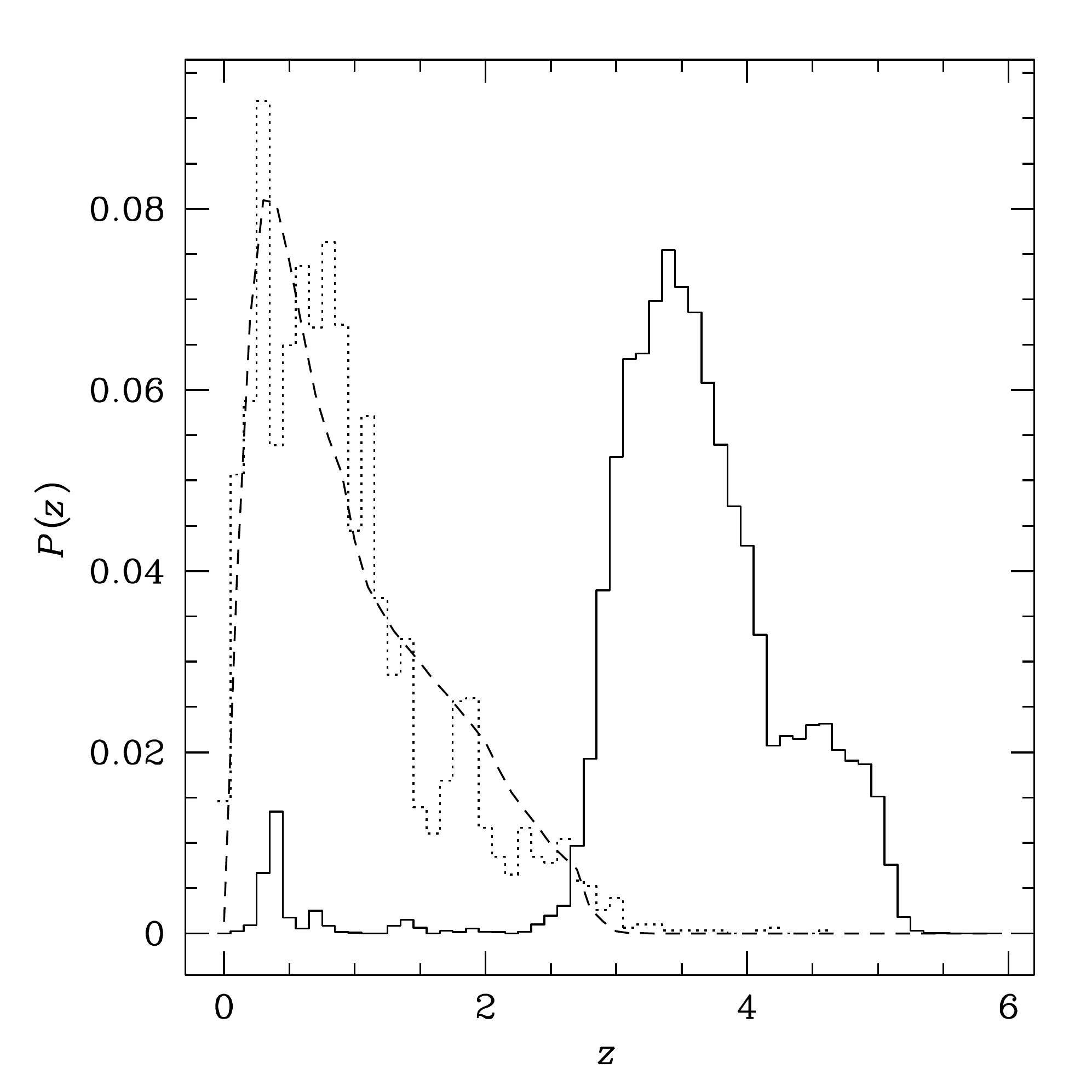}
\caption{Redshift distributions of the sub-millimetre lenses (dashed:
  simulation; dotted: photometric redshifts) and the LBG sources
  (solid) used in the magnification measurement. The separation in
  redshift is not perfect and therefore we model the contribution from
  physical clustering (due to redshift overlap) to the
  cross-correlation signal.}
\label{fig:z_dist}
\end{figure}

The HerMES catalogue contains 3402 objects in the common area of the
optical and sub-millimetre data. For the mass measurement we only
consider lenses with a sub-millimetre flux density at $250\,\mu$m of
$S_{\rm 250}>15\,{\rm mJy}$. This is necessary because confusion (and
other effects) is more severe at fainter flux densities in the
Herschel data, which would make the estimation of a redshift
distribution for the lenses very problematic. This redshift
distribution is needed to interpret the signal, since it depends on
the lens-source geometry, and to estimate a mass. We further apply a
colour cut on the sub-millimetre lenses to enhance the separation in
redshift. We require the ratio of the flux densities at 500\,$\mu$m
and 250\,$\mu$m to be $S_{\rm 500}/S_{\rm 250}<0.5$. This leaves us
with 587 lenses. The measured and simulated redshift distributions of
the lenses are presented in
\cite{2011A&A...529A...4B,2012arXiv1203.1925B}. We use the simulated
one in the following. The model of the redshift distribution of the
sources is taken from \cite{2009A&A...498..725H}. These distributions
are plotted in Fig.~\ref{fig:z_dist}, showing only marginal overlap.

\section{Technique}
\label{sec:technique}
The measurement described here is technically similar to the previous
measurement by \cite{2011MNRAS.414..596W} using sub-millimetre
galaxies and magnification, but conceptually it is very different
because here we are studying the effects of lensing \emph{by}
sub-millimetre galaxies rather than the effects of lensing by other
sources \emph{on} sub-millimetre galaxies \citep[see also][for a
  related cross-correlation measurement]{2006MNRAS.368..732B}. Thus,
we can study how their dark matter halos deflect light and estimate
their average mass through this effect, which was not possible when
the sub-millimetre galaxies were used as sources.

As detailed in other studies
\citep{2010ApJ...723L..13V,2011ApJ...733L..30H} magnification is a
particularly useful tool for measuring the average masses of objects
at high redshifts because -- unlike shear-based weak lensing methods
-- it does not require the background sources to be resolved. A large
fraction of the sub-millimetre galaxies studied here have redshifts
$z>1$. With the best optical ground-based data there are very few
objects at such high redshifts that could be resolved and used as
background sources for shear measurements because of the
seeing-limited size of the optical point-spread-function (PSF). Even
with optical imaging data from the Hubble Space Telescope it becomes
increasingly difficult to resolve objects with $z\ga1.5$ due to PSF
size. Thus, the population of sub-millimetre galaxies studied here
represents an ideal lens sample for magnification when LBGs are used
as background sources.

The WL-magnification effect of the lenses increases the fluxes of
background objects and shifts their positions on the sky. This induces
a change in their number density, which leads to angular correlations
in the positions of lenses and sources on the sky, even in the absence
of physical clustering, i.e., when both populations are well separated
in redshift \citep{2005ApJ...633..589S,2009A&A...507..683H}. Depending
on the slope of the magnitude number counts of the source sample,
$\alpha=2.5\slfrac{{\rm d}\log\left[N(m)\right]}{{\rm d}m}$, a
positive or a negative cross-correlation is expected from WL
theory. Bright galaxies with typically steep slopes ($\alpha-1>0$)
should show positive correlations, and faint galaxies, with shallow
slopes ($\alpha-1<0$), should show anti-correlations.

WL magnification is not the only effect that can cross-correlate the
sky position of objects that are far apart along the line of
sight. Extinction by dust in the foreground objects can also lead to a
characteristic depletion of the number density of background objects,
and this has to be included in the modelling to correctly interpret
the cross-correlation signal. This behaviour can be described by an
effective slope $\alpha_{\rm eff}$ as detailed below.

For this kind of measurement it is particularly important to minimise
the overlap in redshift between the lens and source populations to
suppress a systematic bias due to physical cross-correlations, which
can be much larger than the magnification/extinction signal.

The cross-correlation functions are estimated using the estimator from
\cite{1993ApJ...412...64L} based on pair counts:
\begin{equation}
  w(\theta)=\frac{\rm D_1D_2-D_1R-D_2R}{\rm RR}+1\,,
\end{equation}
with $\rm D_1D_2$ being the number of sub-millimetre-LBG pairs in the
angular range $[\theta,\theta+\delta\theta]$ normalised by the product
of their total numbers, $\rm D_{\rm i}R$ being the normalised number
of pairs between the sub-millimetre/LBG catalogue and a random
catalogue (with the same surface geometry) in that angular range, and
$\rm RR$ being the normalised number of pairs in the random catalogue
in that angular range. By choosing a random catalogue that is much
larger than the data catalogues the shot noise introduced by the
random catalogue is suppressed.

It has been shown in \cite{2003A&A...403..817M} and
\cite{2005ApJ...633..589S} that the signal-to-noise ratio of a
magnification measurement can be boosted when every background galaxy
is given a weight that corresponds to the $\alpha-1$ value at its
magnitude. The same can be done if there is additional dust extinction
by instead using an effective weight, $\alpha_{\rm eff}-1$.

For the optimally-weighted correlation function the same estimator is
used, but $\rm D_1D_2$ and $\rm D_2R$ are replaced by the sums of
weights instead of the number of pairs
\citep{2009A&A...507..683H}. The weights are $\alpha_{\rm eff}-1$ (see
Sect.~\ref{sec:results} for a definition) for the LBG in each pair
accounting for the fact that part of the signal is due to
extinction. Such a weighting maximises the signal-to-noise ratio of
the magnification/extinction measurement
\citep{2003A&A...403..817M,2005ApJ...633..589S}. The gain is
considerable for cases where the weight changes appreciably over the
whole range of magnitudes of the background sample (typically for very
deep data as used here; see also Fig.~\ref{fig:magn_split}).

\section{Theoretical background}
\label{sec:theory}
We assume that the halos hosting the sub-millimetre galaxies can be
described by a single NFW \citep{1996ApJ...462..563N} halo. This model
has two parameters, the mass $M_{200}$ and the concentration
$c$. Detailed dark matter N-body simulations show that these two
parameters are strongly correlated and a relation can be established
between the two. Here, we use the relation by
\cite{2012MNRAS.423.3018P} to reduce the number of parameters in our
mass model. The magnification profile of an NFW halo is described in
\cite{2000ApJ...534...34W}.

We model the signal of the optimally-weighted cross-correlation
function in the following way:
\begin{equation}
w_{\rm opt}\left(\theta\right) = w_\mu\left(\theta\right) + w_\tau\left(\theta\right) + w_{\rm cc}\left(\theta\right)\,,
\label{eq:tot}
\end{equation}
where $w_\mu$ describes the contribution from magnification, $w_\tau$
describes the contribution from dust extinction in the lens galaxies,
and $w_{\rm cc}$ is the physical clustering part of the signal that is
due to redshift overlap between lenses and sources and should be
minimised.

The magnification signal can be calculated from lensing theory:
\begin{equation}
w_\mu\left(\theta\right)=\int_{m_{\rm min}}^{m_{\rm max}}\!\!\!\!\!\!\left[\alpha_{\rm eff}\left(m\right)-1\right]\left[\mu\left(\theta\right)^{\alpha\left(m\right)-1}-1\right]\,\hat N(m)\,{\rm d}m\,,\label{eq:w_mu}
\end{equation}
with $m_{\rm min/max}$ being the faintest and brightest magnitudes of
the background sample, $\alpha_{\rm eff}(m)-1$ being the effective
weight, $\mu(\theta)$ being the magnification profile of an NFW
halo,\footnote{For readability we do not include the redshift
  dependence here. But it should be clear that $\mu(\theta)$ has to be
  calculated by integrating over the redshift distributions of the
  lenses and sources.}  $\alpha(m)-1=2.5\slfrac{{\rm
    d}\log\left[N(m)\right]}{{\rm d}m}-1$ being the weight in absence
of extinction \citep[taken from the LBG luminosity function
of][]{2010A&A...523A..74V}, and $\hat N(m)$ being the normalised
number counts of the LBGs.

The dust absorption $A$ is related to the optical depth through
\begin{equation}
A=\slfrac{2.5}{\ln(10)}\tau=1.08\tau\,.
\end{equation}
We assume that $A$ and the magnification excess $\delta\mu=\mu-1$ are
related by\footnote{Note that this does not constrain the dust profile
  yet because $c_{\rm d}$ can be a function of angular scale.}
\begin{equation}
A=c_{\rm d}\delta\mu\,.
\end{equation}
Here we assume that this is true on average. The redshift dependence
of the dust extinction and the magnification is certainly
different. Hence halos at different redshift will contribute
differently to the magnification and extinction signals. We neglect
this effect in the following but note that firm conclusions on the
shape of the dust profile can only be drawn if such effects are
included in the modelling. The mass estimate is not directly affected
by this simplification. Under these assumptions the dust signal
becomes:
\begin{align}
w_\tau\left(\theta\right)=&\int_{m_{\rm min}}^{m_{\rm max}}\!\!\!\!\!\!\left[\alpha_{\rm eff}\left(m\right)-1\right]\nonumber\\
&\times\left[1.08\:c_{\rm d}10^{-0.4\alpha\left(m\right)}\:\delta\mu\left(\theta\right)-1\right]\hat N(m)\, {\rm d}m\,.\label{eq:w_tau}
\end{align}

The contribution from physical clustering to the angular correlation
function are modelled by
\begin{align}
w_{\rm cc}(\theta)=&\int_{m_{\rm min}}^{m_{\rm max}}\!\!\!\!\!\!\left[\alpha_{\rm eff}\left(m\right)-1\right]b_1b_2w_{\rm DM}(\theta)\hat N(m)\, {\rm d}m\,,\label{eq:w_cc}
\end{align}
where $b_{1/2}$ are the average bias factors of the lenses and
sources, respectively, and $w_{\rm DM}$ is the angular correlation
function of the dark matter field\footnote{This is calculated with the
  equation from \cite{1953ApJ...117..134L}. We used the code by
  \cite{2004MNRAS.347..813H} for this. The contribution from physical
  clustering has to be weighed appropriately with the product of the
  redshift distributions of lenses and sources.}. We conservatively
estimate $b_1=b_2=2$ in our analysis.

\section{Results}
\label{sec:results}
\begin{figure}
\includegraphics[width=\hsize]{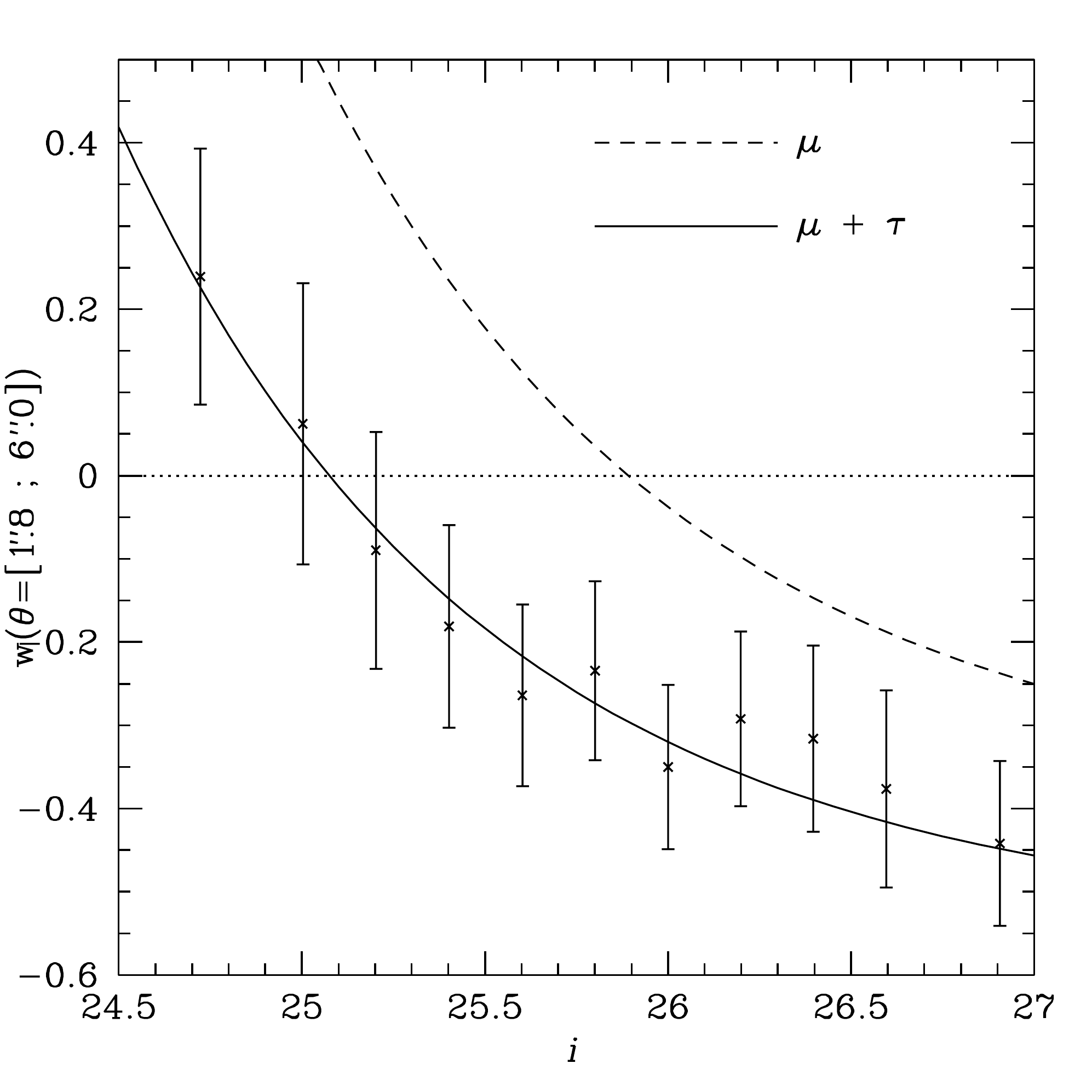}
\caption{Amplitude of the angular cross-correlation function between
  3402 sub-millimetre lenses and LBG sources as a function of the LBG
  $i$-band magnitude in the angular interval
  $1\farcs8<\theta<6\farcs0$. Errors are Poissonian. The dashed line
  represents the expected amplitude if there was only
  magnification. The solid line represents a more realistic model
  including the effects of extinction by dust in the lenses.}
\label{fig:magn_split}
\end{figure}

The angular cross-correlation (unweighted) between the positions of
all sub-millimetre lenses in the 1\,deg$^2$ area (3402 objects; no
flux cut) and the LBG sources in different magnitude slices at close
separations is shown in Fig.~\ref{fig:magn_split}. Here we use one
broad angular bin with $1\farcs 8<\theta<6\farcs 0$. As expected, the
positions of the brighter background galaxies are positively
correlated with the positions of the lenses, while the positions of
the fainter ones are anti-correlated. However, the dashed line, which
shows the expected amplitude from magnification only, without dust
extinction, does not fit the data points. The solid line represents
the expected amplitude of the cross-correlation function assuming that
on average the dust absorption is proportional to the magnification
excess \citep[note that this is well supported by results from][using
SDSS data]{2010MNRAS.405.1025M}. We fit for the dust normalisation
constant $c_{\rm d}$, obtaining a best-fit value of $c_{\rm
  d}=0.35$. The overall normalisation was left as another free
parameter and both predictions were multiplied by this value.  In both
cases we model $N(m)$ from the LBG luminosity function estimates by
\cite{2010A&A...523A..74V}.

\cite{2010MNRAS.405.1025M} found a value of $c_{\rm d}\approx0.1$
using a magnitude-limited, optically-selected, low-$z$ galaxy sample
from the Sloan Digital Sky Survey. Thus, the sub-millimetre galaxies
we study here are -- not surprisingly -- considerably more dusty
(although the $c_{\rm d}$ values cannot be compared directly because
of the different rest frame wavelengths of the filters used for the
selection of the lenses).

Physical clustering signals due to redshift overlap between lenses and
sources are always positive. Thus, detecting a negative
cross-correlation, as shown in Fig.~\ref{fig:magn_split}, already
suggests that we do not have significant redshift overlap
here.

Here, we use all 3402 lenses detected in the sub-millimetre data to
improve the statistics and be able to constrain the scaling between
magnification and extinction. However, we limit the following analysis
to the 587 objects with $S_{\rm 250}>15\,{\rm mJy}$ and $S_{\rm
  500}/S_{\rm 250}<0.5$, for which we can estimate a reliable redshift
distribution (see Sect.~\ref{sec:z-dist}). By applying the $c_{\rm d}$
value found from the whole sample of 3402 objects to the
high-confidence sample of 587 objects we implicitly assume that the
redshift distribution, as well as the scaling between $\delta \mu$ and
$A$, does not differ significantly between the two samples. With the
current data and models this cannot be tested directly and has to
remain an assumption here.

\begin{figure}
\includegraphics[width=\hsize]{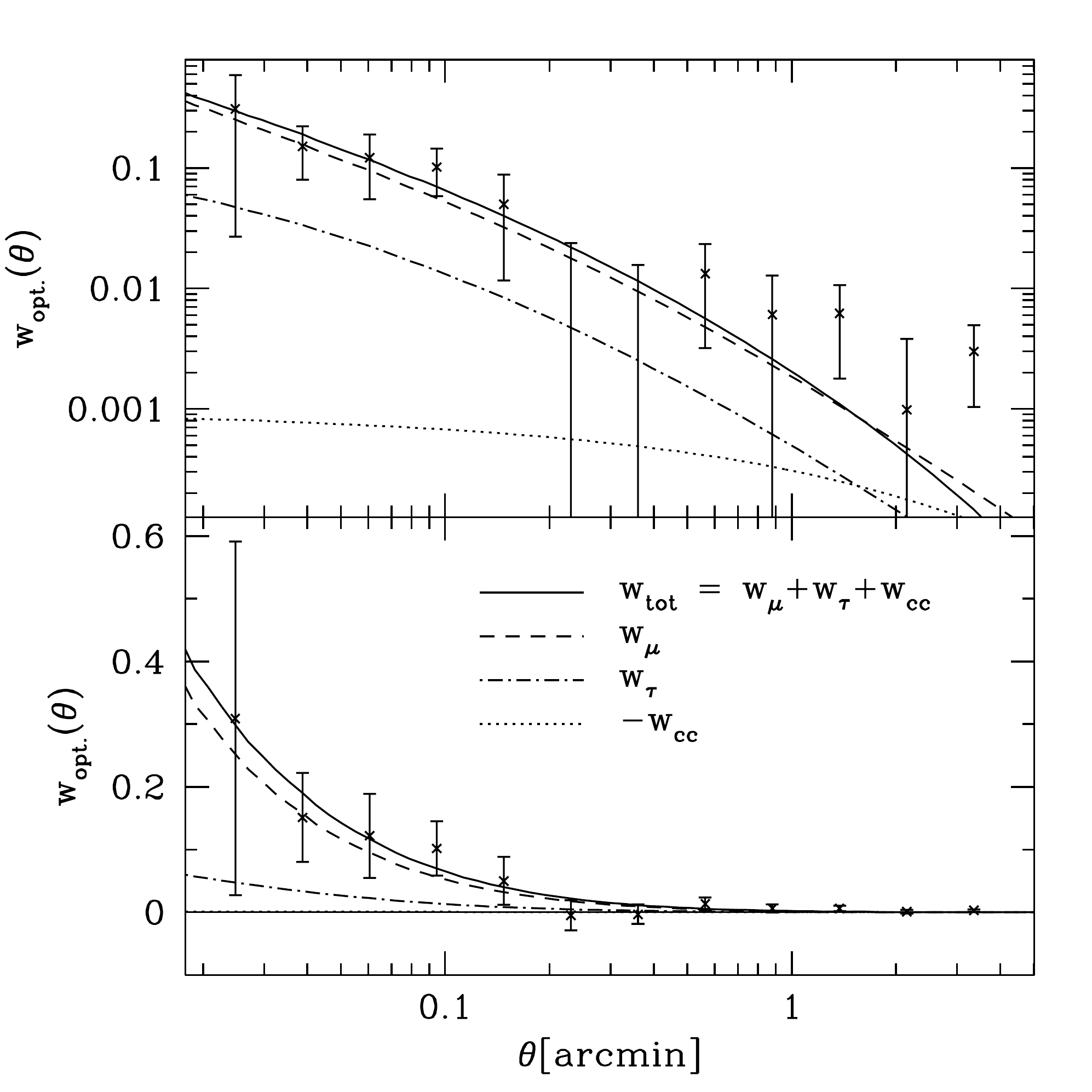}
\caption{Optimally-weighted angular correlation function between 587
  sub-millimetre galaxies with 250\,$\mu$m flux density
  $S_{250}>15\,{\rm mJy}$ and all background LBGs. The weights are
  based on the results presented in Fig.~\ref{fig:magn_split} with
  only the brightest Lyman-break galaxies getting a positive weight
  and the fainter ones all getting a negative weight. The solid line
  represents the best-fit model, consisting of contributions from
  magnification (dashed line), dust extinction (dot-dashed), and a
  negative contribution from physical clustering between lenses and
  sources (dotted, and almost negligible). The top panel shows a
  logarithmic scale for the correlation amplitude whereas the bottom
  panel has a linear scale. Errors are estimated from a jack-knife
  resampling of the background population.}
\label{fig:magn_opt}
\end{figure}

In Fig.~\ref{fig:magn_opt} the optimally-weighted cross-correlation
function between these 587 sub-millimetre galaxies and the full set of
background LBGs is shown, with errors estimated from a jack-knife
resampling of the background sample. We detect a signal at the
7-$\sigma$ level. The best-fit value for $c_{\rm d}$ ($=0.35$) that we
take from the previous measurement indicates that the majority of this
signal is due to magnification and not extinction. Note that the
accurate relative contributions depend on the absolute value of the
magnification itself (and hence also on angular scale). At small
angular scales magnification contributes about 85\% to the signal,
whereas at large angular scales its contribution drops to about 75\%.

We fit the model from Eq.~\ref{eq:tot} to the data, which represents
the magnification of an NFW dark matter halo, as well as the
extinction of a dust halo whose dust is distributed according to the
same profile.  Additionally we take the small contribution from
physical cross-correlations $w_{\rm cc}$ into account, which is
further suppressed because of the weighting scheme. Due to the fact
that most of the background LBGs have negative weights, $\alpha_{\rm
  eff}-1$, $w_{\rm cc}$ actually becomes negative in this
optimally-weighted case.\footnote{ This also means that not accounting
  for or underestimating $w_{\rm cc}$ would lead to an underestimate
  of the mass.} We also account for errors in the centroid positions
of the sub-millimetre lenses due to the limited resolution of
Herschel.

The best-fit estimate of the average NFW halo mass within $r_{200}$ is
$\log_{10} \left[M_{200}/ M_\odot\right] = 13.17^{+0.05}_{-0.08}{\rm
  (stat.)}$ or
$M_{200}=1.48^{+0.18}_{-0.25}{\rm(stat.)}\times10^{13}M_\odot$,
directly confirming that sub-millimetre galaxies are hosted by very
massive halos.

Note that by assuming that dust follows mass (see above) we can
exploit the contribution from extinction to the signal in order to
actually constrain the mass. The quoted error contains contributions
from shot-noise as well as from clustering of the background
population. No uncertainties originating from the measurement of
$c_{\rm d}$, the assumption that dust follows mass, the assumed
redshift distributions, or the physical clustering model are included
here, because these uncertainties are very hard to quantify at the
current stage, but could well be substantial.

Our best-fit model also suggests that sub-millimetre galaxies are very
dusty. Assuming that the extinction law of the dust in these galaxies
is similar to that found in the Small Magellanic Cloud
\citep{2010MNRAS.405.1025M}, the dust mass fraction at the scales we
are probing ($\ga10\,$kpc) can be estimated to be roughly $M_{\rm
  dust}/M_{200}\approx6\times10^{-5}$.

\section{Discussion}
\label{sec:discussion}
\subsection{Details of the modelling}
The magnification exceeds $\mu=1.5$ for the smallest scales that we
are probing here, taking into account that the dust extinction
contributes 15\% of this signal. Thus, this is not strictly weak
lensing but rather an intermediate regime between strong and weak
lensing where a statistical technique -- similar to what is used in
weak lensing -- is applied. Note that because of this we do not assume
the weak lensing limit (i.e. $\delta\mu\ll1$) in any part of the
modelling.

Both, the magnification as well as the extinction, lead to a net
depletion of background sources behind the sub-millimetre lenses,
hence the negative average weight. Even without directly using colour
information, we can still disentangle the two effects because we have
a measurement of the average intrinsic luminosity function of the LBGs
\citep{2010A&A...523A..74V}. Attributing the amplitude of the
correlation function solely to either magnification or extinction
would lead to inconsistencies as it is shown in
Fig.~\ref{fig:magn_split}.

\subsection{Relation to Clustering}
Clustering measurements have been used to constrain the dark matter
halo masses of sub-millimetre galaxies by interpreting their
auto-correlation function in the framework of the halo model
\citep{2006ApJ...641L..17F,2010A&A...518L..22C,2011Natur.470..510A,2012MNRAS.421..284H,2012arXiv1208.5049V}.
While we also measure a correlation function here, the origin of the
signal is very different. The lensing magnification method described
here differs in some important aspects from such clustering
measurements.

We actually try to suppress the physical clustering contribution to
the cross-correlation function as much as possible by separating the
sub-millimetre lenses and LBG sources in redshift. The
auto-correlation signal that is used by physical clustering
measurements represents a systematic nuisance in magnification
studies. We model its small contribution to the cross-correlation
function here and account for it in the mass measurement, but it is
important to note that it is not used to estimate the mass.

In order to interpret an auto-correlation signal from physical
clustering, some model for the galaxy bias has to be assumed. No such
model is needed to interpret our lensing correlation function on small
scales, under the assumption that all sub-millimetre galaxies are
central galaxies to their halos.

Another important difference between physical clustering measurements
and lensing magnification is the different sensitivity to the shape of
the redshift distribution. While the interpretation of the physical
angular cross-correlation function depends critically on the width of
the redshift distribution, lensing is fairly insensitive to this width
because the lensing efficiency is a slowly changing function of
redshift. It is mainly the mean redshift that is important
here. However, this is only true as long as there is no redshift
overlap between lenses and sources. Once there are physical
cross-correlations contributing to the signal, as in the measurement
presented here, the width becomes important again for estimating this
contribution.

\subsection{Comparison to abundance matching} 
We also roughly estimate the mass of halos hosting the sub-millimetre
galaxies from abundance matching \citep{2012A&A...537L...5B} using the
following method. We compute the infrared luminosity of a source
having the mean redshift and the mean flux of the Herschel sample. We
convert this luminosity into star formation rate using the
\cite{1998ApJ...498..541K} constant ($1\times10^{-10}\,M_\odot/{\rm
  yr}/L_\odot$). We finally use the abundance matching results at this
SFR and this mean redshift, using an interpolation between the two
closest redshifts where the abundance matching was performed. This
yields a mass estimate of $\log_{10} \left[M_{200}/ M_\odot\right] =
13.2\pm0.2$, in very good agreement with our magnification/extinction
measurement.\footnote{It should be noted that the estimate from IR
  abundance matching is associated with a number of systematic errors
  that are not present in the magnification measurement.}

\subsection{Notes on the sub-millimetre catalogue}
This sub-millimetre galaxy lens sample represents a fairly faint
population. Thus, the number of lenses that are lensed themselves is
also very low. A completely negligible fraction of the signal shown in
Fig.~\ref{fig:magn_opt} is due to lensing of both, the lenses and the
sources, by foreground structures in front of both
\citep{2011MNRAS.415.1681H}.

When we use the full catalogue of 3402 objects for the determination
of the constant $c_{\rm d}$ it is very probable that there are a large
number of spurious objects in this sample. Note that this does not
matter in that particular case because we are only interested in the
relative amounts of magnification and extinction there. A spurious
source should not add to either of the two and hence not change the
result. For the actual mass measurement we use only galaxies with
250\,$\mu$m flux densities of $S_{\rm 250}>15\,{\rm mJy}$ as
lenses. Given the depth of the data this sample should not contain any
spurious sources.

\subsection{Limitations and possible extensions}
There are several aspects that limit the accuracy of the measurement
presented here. We have to use a redshift distribution based on a
model for the sub-millimetre lenses to predict the magnification
signal. Individual (photometric) redshifts for all sub-millimetre
galaxies would certainly help to overcome any uncertainties in the
modelling of their redshift distribution \citep{2012arXiv1203.1925B}.

The amount of physical cross-correlation, described by $w_{\rm cc}$,
depends on the redshift distribution of the LBGs, which is taken from
simulations \citep{2009A&A...498..725H}. It should be noted that the
choice here is actually quite pessimistic. With different plausible
choices presented in \cite{2009A&A...498..725H}, which are based on
modified simulations, the overlap and hence amplitude of $w_{\rm cc}$
would decrease even further. The mass estimate is virtually unaffected
by this very small change. We would like to stress that the optimal
weighting of the correlation function greatly suppresses the physical
cross-correlation. In order to boost the signal-to-noise ratio of the
desired signal components each background galaxy is weighted with its
expected responsiveness to the combined effects of magnification and
extinction. These weights are estimated from the background galaxies'
magnitudes. However, the scaling of the physical cross-correlation
signal with magnitude of the background galaxies is completely
different. This suppression is reflected in the very low amplitude of
$w_{\rm cc}$ compared to $w_\mu$ and $w_\tau$ in
Fig.~\ref{fig:magn_opt}.

We assume that the optical depth of dust extinction in the lenses,
$\tau$, follows the magnification excess, $\delta\mu$, such that
$\frac{2.5}{\ln(10)}\tau=c_{\rm d}\delta\mu$. This is motivated by the
measurement by \cite{2010MNRAS.405.1025M} with low-$z$ galaxies but
has not been shown to be true for sub-millimetre galaxies. We can only
constrain the value of $c_{\rm d}$ at small scales where the
signal-to-noise ratio is large enough to split up the LBG background
sample into several magnitude bins: hence, the choice of the broad
angular bin with $1\farcs8<\theta<6\farcs0$ in
Fig.~\ref{fig:magn_split}. However, our measurement over 1\,deg$^2$ is
not powerful enough to fully constrain the angular dependence of the
extinction. Future measurements over larger areas could exploit the
reddening effect \citep{2010MNRAS.405.1025M} to constrain this
dependence and potentially even the extinction law of the dust in
sub-millimetre galaxies.

For consistency, we also calculate models where the dust is assumed to
be distributed according to an exponential profile with different
scale lengths $h_{\rm R}=5-40$\,kpc instead of an NFW-like dust
halo. The factor $c_{\rm d}$ becomes a function of angular scale then,
and so do the weights for the optimally weighted correlation
function. The overall amplitude of these dust models is fixed by the
measured, integrated value of $c_{\rm d}$ at small scales (see
Fig.~\ref{fig:magn_split}). The best-fit mass for such models is
smaller by a factor 2--3 than the best-fit mass for the model with an
NFW-like dust halo -- the larger the scale length the smaller the
discrepancy. However, the reduced $\chi^2$ of these model fits to the
data, $\chi^2/{\rm dof}=2$--$5$, is considerably larger than the
fiducial $\chi^2/{\rm dof}=1.5$ of the model with an NFW-like dust
component -- again with the largest dust scale length (40\,kpc)
yielding the smallest $\chi^2/{\rm dof}$. This indicates that the dust
is indeed very widely distributed, and our measurements favour the
fiducial model where ``dust follows mass''. It should be stressed that
we do not interpret the results in such a way that these
sub-millimetre galaxies have a smooth NFW-like dust halo or
exponential dust disks with very large scale-lengths of $h_{\rm
  R}\ga40$\,kpc. It is, however, possible that some additional, widely
extended dust component -- similar to what is found in
\cite{2010MNRAS.405.1025M} -- is responsible for the effect seen here.

Another consequence of the limited signal-to-noise ratio is that we
cannot fit a multi-parameter model to the data points in
Fig.\ref{fig:magn_opt}. For this reason we concentrate on small
angular scales, where a single halo can be assumed to dominate (1-halo
term). The contributions from other halos (2-halo term) only become
important at larger angular scales and can be neglected here for
simplicity. Further assuming a mass-concentration relation
\citep{2012MNRAS.423.3018P} for the NFW halo leaves us with just one
parameter to fit, the average mass $M_{200}$. We check the influence
of our choice of a particular mass-concentration relation by also
implementing the relation by \cite{2011MNRAS.411..584M} and find a
negligible difference in the best-fit mass of $\Delta\log_{10}
\left[M_{200}/ M_\odot\right]\approx0.03$. Again, the interpretation
of a measurement from a more powerful, larger-area survey could easily
be extended to constrain the halo-occupation distributions, satellite
fractions, and concentration parameters directly.

We also assume that all sub-millimetre galaxies are central galaxies
to their halos. There are some indications in the literature from
clustering studies that a fair fraction of the galaxies in such a
sub-millimetre sample are in fact satellites \citep[$\sim25\%$ for our
lens sample; see][]{2010A&A...518L..22C}. In the regime probed here
the magnification excess, $\delta_\mu$, is fairly linear in the
mass. Thus, assuming a worst-case scenario, where no mass would be
associated with 25\% of our lenses, this would lead to an
underestimate of the halo mass of the remaining 75\% central galaxies
of $\Delta\log_{10} \left[M_{200}/ M_\odot\right]\approx0.1$. This is
similar to the statistical error of our measurement so that we decided
not to correct for this directly here, especially because the real
effect is certainly smaller than this worst-case scenario.

\section{Summary and outlook}
\label{sec:outlook}
In this paper we show how to use the magnification effect of WL to
measure the average mass of sub-millimetre galaxies. Using
sub-millimetre galaxies as the lenses and Lyman-break galaxies as the
sources we find a mass of $\log_{10} \left[M_{200}/ M_\odot\right] =
13.17^{+0.05}_{-0.08}{\rm (stat.)}$ for the halos hosting the
sub-millimetre galaxies. The presence of significant amounts of
extinction by dust in the lenses complicates this
measurement. However, accounting for the dust allows us to
simultaneously constrain the dust mass fraction of the lenses to
$M_{\rm dust}/M_{200}\approx6\times10^{-5}$.

With deep, large-area, imaging surveys (DES, LSST, Euclid) on the
horizon, WL-magnification methods will gain additional importance. The
higher the redshifts of the objects under study, the fewer the
techniques that can provide a reliable mass measurement. As we have
shown in this paper, magnification can provide such mass estimates for
high redshift objects. Follow-up observations of the optical surveys
mentioned above with sub-millimetre telescopes will provide very large
samples of high-$z$ star-forming galaxies. Refined magnification
measurements of the kind presented here, alongside with clustering
measurements and other techniques, will yield unprecedented insights
into the physical processes of star-formation in the first half of the
Universe.

Future measurements with better statistics will enable the lensing and
extinction effects to be separated. Additional redshift information
for the sub-millimetre galaxies will lead to more accurate mass
estimates. By splitting the sub-millimetre galaxy sample in brightness
it will be possible to directly study the relationship between mass
and star-formation rate.

This study shows that extinction can not be neglected in magnification
studies. The studies by \cite{2009A&A...507..683H} and
\cite{2012MNRAS.426.2489M} show excessive anti-correlations when very
faint sources are used. Also the amplitudes of the angular correlation
function presented there turn over from positive to negative at
brighter source magnitudes than expected from magnification
alone. This hints at extinction playing a role. The framework outlined
here can explain this effect, and we strongly suggest future
magnification studies to account for extinction to avoid a systematic
bias.

\section*{Acknowledgements}
Herschel is an ESA space observatory with science instruments provided
by European-led Principal Investigator consortia and with important
participation from NASA.\\SPIRE has been developed by a consortium of
institutes led by Cardiff University (UK) and including:
Univ.\ Lethbridge (Canada); NAOC (China); CEA, LAM (France); IFSI,
Univ.\ Padua (Italy); IAC (Spain); Stockholm Observatory (Sweden);
Imperial College London, RAL, UCL-MSSL, UKATC, Univ.\ Sussex (UK); and
Caltech/JPL, IPAC, Univ.\ Colorado (USA). This development has been
supported by national funding agencies: CSA (Canada); NAOC (China);
CEA, CNES, CNRS (France); ASI (Italy); MCINN (Spain); SNSB (Sweden);
STFC, UKSA (UK); and NASA (USA),\\ This research has made use of data
from the HerMES project (\url{http://hermes.sussex.ac.uk/}). HerMES is
a Herschel Key Programme utilising Guaranteed Time from the SPIRE
instrument team, ESAC scientists and a mission scientist.\\ The HerMES
data were accessed through the HeDaM data-base
(\url{http://hedam.oamp.fr}) operated by CeSAM and hosted by the
Laboratoire d'Astrophysique de Marseille.\\ The optical data are based
on observations obtained with MegaPrime/MegaCam, a joint project of
CFHT and CEA/DAPNIA, at the Canada-France-Hawaii Telescope (CFHT)
which is operated by the National Research Council (NRC) of Canada,
the Institut National des Sciences de l'Univers of the Centre National
de la Recherche Scientifique (CNRS) of France, and the University of
Hawaii. This work is based in part on data products produced at
TERAPIX and the Canadian Astronomy Data Centre, as part of the
Canada-France-Hawaii Telescope Legacy Survey, a collaborative project
of NRC and CNRS.\\ HH is supported by the Marie Curie IOF 252760, a
CITA National Fellowship, and the DFG grant Hi 1495/2-1. LVW is
supported by NSERC and CIfAR. DS acknowledges support by NSERC and
CSA. TE is supported by the Deutsche Forschungsgemeinschaft through
project ER 327/3-1 and the Transregional Collaborative Research Centre
TR 33 - ”The Dark Universe”. RVDB acknowledges support from the
Netherlands Organisation for Scientic Research grant number
639.042.814.

\small
\textbf{Author Contributions:} 

H.H. led the analysis and wrote the draft version of the
paper. L.v.W. and D.S. contributed significantly in the development of
this project through ideas and discussions on a daily
basis. M.B. provided the redshift distribution of the sub-millimetre
galaxies. T.E. led the optical data reduction and
calibration. R.F.J.v.d.B. provided the luminosity function estimate of
the background galaxies. All other co-authors contributed extensively
and equally by their varied contributions to the SPIRE instrument, the
Herschel mission, analysis of SPIRE and HerMES data, planning of
HerMES observations and scientific support of HerMES, and by
commenting on this manuscript as part of an internal review process.

\normalsize

\bibliographystyle{mn2e_mod}

\bibliography{D2_submm_magn}

\label{lastpage}

\end{document}